\documentclass[twocolumn,aps,showpacs,prb,tightenlines,amsmath,amssymb,superscriptaddress]{revtex4}
\usepackage{graphicx}

\usepackage{dcolumn}

\usepackage{bm}

\usepackage{colordvi}
\usepackage{color}
\begin{document}

\title{Cyclotron effect on coherent spin precession of two-dimensional electrons}

\author{M. Griesbeck}
\affiliation{Institut f\"ur Experimentelle und Angewandte Physik,
Universit\"at Regensburg, D-93040 Regensburg, Germany}
\author{M.M.\ Glazov}
\email{glazov@coherent.ioffe.ru}
\affiliation{A. F. Ioffe Physical-Technical Institute, Russian Academy of Sciences, 194021 St. Petersburg, Russia}
\author{T.\ Korn}
\email{tobias.korn@physik.uni-r.de}
\affiliation{Institut f\"ur Experimentelle und Angewandte Physik,
Universit\"at Regensburg, D-93040 Regensburg, Germany}
\author{E.Ya.\ Sherman}
\affiliation{Department of Physical Chemistry, The University of the Basque Country, 48080 Bilbao, Spain \\ IKERBASQUE Basque Foundation for Science, Alameda Urquijo 36-5, 48011, Bilbao, Bizkaia, Spain}
\author{D.~Waller}
\affiliation{Institut f\"ur Experimentelle und Angewandte Physik,
Universit\"at Regensburg, D-93040 Regensburg, Germany}
\author{C. Reichl}
\affiliation{Institut f\"ur Experimentelle und Angewandte Physik,
Universit\"at Regensburg, D-93040 Regensburg, Germany}
\author{D.\ Schuh}
\affiliation{Institut f\"ur Experimentelle und Angewandte Physik,
Universit\"at Regensburg, D-93040 Regensburg, Germany}
\author{W.\ Wegscheider\footnote{present address: Solid State Physics Laboratory, ETH Zurich, 8093 Zurich, Switzerland}}
\affiliation{Institut f\"ur Experimentelle und Angewandte Physik,
Universit\"at Regensburg, D-93040 Regensburg, Germany}
\author{C.\ Sch\"uller}
\affiliation{Institut f\"ur Experimentelle und Angewandte Physik,
Universit\"at Regensburg, D-93040 Regensburg, Germany}

\date{\today}

\begin{abstract}
We investigate the spin dynamics of  high-mobility two-dimensional electrons in  GaAs/AlGaAs quantum wells grown along the $[001]$ and $[110]$ directions by time-resolved Faraday rotation at low temperatures. In measurements on the $(001)$-grown structures without external magnetic fields, we observe coherent oscillations of the electron spin polarization about the effective spin-orbit field. In non-quantizing magnetic fields applied normal to the sample plane, the cyclotron motion of the electrons rotates the effective spin-orbit field. This rotation leads to fast oscillations in the spin polarization about a non-zero value and a strong increase in the spin dephasing time in our experiments. These two effects are absent in the $(110)$-grown structure due to the different symmetry of its effective spin-orbit field. The measurements are in excellent agreement with our theoretical model.
\end{abstract}


\maketitle

A key issue in the semiconductor spintronics~\cite{Fabian,Fabian07} is the dynamics of spins
of carriers in low-dimensional structures. Advances in the technology allow manufacture
of very clean two-dimensional electron systems (2DES) based on GaAs/AlGaAs heterostructures where
carriers can move ballistically over distances of several micrometers and the time between
scattering events is on the order of 100~ps. The main spin dephasing mechanism in these structures is the Dyakonov-Perel' (DP) mechanism~\cite{dp}, driven by the wave vector $\bm k$-dependent spin-orbit (SO) fields $\bm \Omega_{\bm k}$ present due to bulk inversion asymmetry (Dresselhaus field)~\cite{bia} and structure inversion asymmetry (Rashba field)\cite{sia}. By growing 2DES along different crystallographic directions, the symmetry of the SO fields can be changed, leading to strong modifications of the electron spin dynamics: in [110]-grown 2DES, for example, the Dresselhaus field points along the growth direction, regardless of the electron wave vector. Electron spins aligned along the growth direction  experience no torque, and the DP mechanism is therefore effectively suppressed for this spin orientation \cite{dp1,Ohno}, while it is still the main spin dephasing channel for any other spin orientation, leading to a strong orientational anisotropy of the spin dephasing time (SDT).~\cite{wugo,Dohrmann} In the presence of an additional Rashba field caused by an asymmetrical placement of the modulation doping layers~\cite{Belkov_PRL,Olbrich_PRB} or tuned by an external gate voltage~\cite{Karimov}, and oriented along the 2DES plane regardless of growth direction, the SDT is reduced drastically.
By contrast, in [001]-grown 2DES Dresselhaus and Rashba fields are in-plane, hence, the spin $z$-component dephasing is relatively fast, and their interference may also lead to a strong orientational anisotropy of spin dephasing.~\cite{Averkiev,Averkiev06}

Two regimes of the DP mechanism are typically considered. The first is
the `motional narrowing regime', where the electron spin precesses about the SO field only by a small angle in between  scattering events. The other one is the `weak scattering regime' where the spin can precess one or more full revolutions before the electron is scattered.
At liquid-helium temperatures and above, most 2DES are in the motional narrowing regime. In high-mobility samples at low temperatures, however, the weak scattering regime becomes accessible in the experiment and the precession of the electron spins about the internal SO fields is observable as coherent oscillations of the $z$-component of the spin polarization.~\cite{Brand,Leyland07,stich} Unlike the `motional narrowing regime' where the cyclotron effect of an external magnetic field simply results in a spin relaxation slow-down~\cite{Ivchenko73}, in high-mobility systems the regular change in the electron wave vector caused by the cyclotron rotation was predicted to have a dramatic effect on spin dynamics.~\cite{Glazov_SSC,Glazov_EPL}

Here, we present a study on the electron spin dynamics in  high-mobility 2DES  at low temperatures in zero and non-quantizing perpendicular fields. We observe coherent zero-field oscillations, and fast, small-amplitude spin beats in perpendicular fields which stem from the cyclotron rotation of the SO field. Our samples were grown by molecular beam epitaxy on $(001)$ and $(110)$-oriented semi-insulating GaAs substrates. The active region in all three samples is a GaAs-Al$_{0.3}$Ga$_{0.7}$As single quantum well (QW). While sample B is a conventional, one-side modulation-doped QW, samples A and C utilize a complex growth structure (similar to that used in Ref.~\onlinecite{Umansky}) to achieve a symmetrical, double-sided doping profile.  Sample properties are listed in Table~\ref{Data}. For optical measurements in transmission geometry, the samples were glued onto sapphire substrates with an optical adhesive, and the substrate and buffer layers were removed by selective
etching.

\begin{table}
\caption{{\bf Sample data.} Density and mobility were determined
from magnetotransport measurements at 1.3~K. The momentum scattering time $\tau_p$ is found from the mobility. The electron-electron scattering time $\tau^*_{ee}$ is calculated for $4.5$~K.~\cite{gi}} \label{Data}
\begin{tabular}{|c|c|c|c|c|c|c|}
  \hline
  \#& growth  & width  & density $n$  &  mobility $\mu$ & $\tau_p$ & $\tau^*_{ee}$ \\
      & axis  &(nm) & ($10^{11}$cm$^{-2}$) & ($10^{6}$~cm$^2/$Vs)&  (ps) & (ps)\\
  \hline
  A &  (001) & 30 & 2.97 & 14.8 & 563& 88\\ 
  B & (001) & 20 &2.1 & 1.6  & 61 & 22\\ 
  C &  (110) & 30 & 3.4 & 5.1 & 194 & 130\\    
   \hline
\end{tabular}
\end{table}

In the time-resolved Faraday rotation (TRFR) measurements, a
mode-locked Ti:Sapphire laser  was used, which allowed for a near-resonant excitation of electrons into the conduction band slightly above the Fermi energy of the 2DES. Details of the experimental setup are published elsewhere.~\cite{Stich_PRB}  The TRFR measurements were performed in a split-coil magnet cryostat with a $^3$He insert, allowing for sample temperatures between 400~mK and 4.5~K.

\begin{figure}
  \includegraphics[width= 0.4\textwidth]{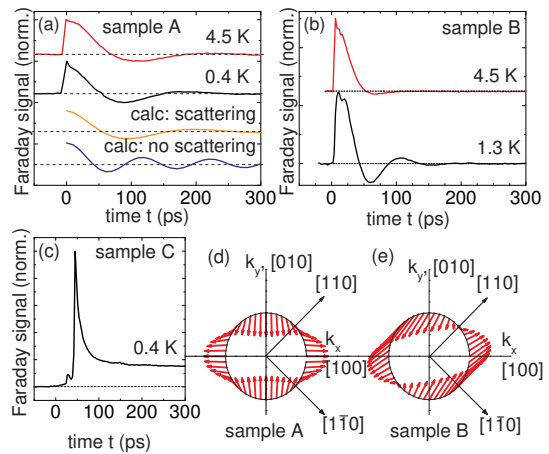}
   \caption{{\bf (a)-(c): TRFR measurements without external magnetic field}. (a) Sample A, measured at 4.5~K and 400~mK. Calculated data~\cite{model_param} with and without the inclucion of scattering (as marked near the lines)  are shown below the measured data.
   (b) Sample B, measured at 4.5~K and 1.3~K. Sample C, measured at 0.4~K. (d) Symmetry of the SO field of sample A for electron spins at the Fermi surface. The arrows indicate the direction and amplitude of the SO field. (e) Symmetry of the SO field of sample B.}
   \label{Zerofield}
\end{figure}

First, we study the spin dynamics in all three samples in the absence of an external magnetic field. Figures \ref{Zerofield}(a)-(c) show TRFR traces measured on samples A-C at low temperatures.  In samples A and B, a strongly damped oscillation is clearly observable in the measurements at 4.5~K. In both samples, the decay constant of this oscillation becomes longer, and the oscillation frequency increases, as the sample temperature is lowered. In contrast, sample C does not show oscillatory spin dynamics, even at low temperatures. Instead, the TRFR signal decays partially within the measurement window and then approaches a nonzero value.

In samples A and B the SO interaction-induced effective magnetic field $\bm \Omega_{\bm k}$ lies in the QW plane. In the weak scattering regime $\Omega_{\bm k_{\rm F}}\tau^*\gg 1$  the electron spin precesses around $\bm \Omega_{\bm k}$ and the pronounced oscillations are observed. Here, $\Omega_{\bm k_{\rm F}}$ is the effective magnetic field for electrons at the Fermi level and $\tau^*$ is the single electron momentum scattering time which has additive contributions from disorder and phonon scattering processes,  as well as from the electron-electron scattering.~\cite{gi} The damping of the oscillations is caused by these scattering processes and by the anisotropy of the spin splitting.~\cite{Glazov_SSC} In the symmetric sample A, the Dresselhaus term contains first and third angular harmonics as is schematically shown in Fig.~\ref{Zerofield}(d) since for the given sample parameters, the Fermi wave vector and the inverse quantum well width are comparable. The spin beats are damped due to the spin splitting anisotropy and electron-electron scattering: it is seen that the change in sample temperature from 400~mK to 4.5~K (Fig. \ref{Zerofield}(a)), which reduces the electron mobility by a factor of two, only weakly reduces the decay constant of the coherent oscillation from 74~ps at 400~mK to 63~ps at 4.5~K. As the calculated traces in Fig.~\ref{Zerofield}(a) show, the spin splitting anisotropy alone is not sufficient to explain the damping, therefore, electron-electron scattering is likely to be the dominant scattering process in this sample also at 400~mK. The weak temperature effect on the beats may be attributed to the heating of the electron gas by the excitation which leads to a reduced difference in the electron-electron scattering times at the two measurement temperatures.

 In sample B, which has a one-side doping layer, Rashba and Dresselhaus fields are present; their relative strength was determined by magnetoanisotropy measurements~\cite{Stich_PRBRC07} to be about $0.65$. The symmetry of their vector sum is shown in Fig. \ref{Zerofield}(e). The average amplitude of this effective SO field is larger than for sample A, due to the Rashba effect and the enhancement of the Dresselhaus term for the thinner QW.~\cite{Leyland07} Accordingly, the precession frequency of the coherent beats is larger. In this sample, which has a lower mobility than sample A, the coherent oscillation is predominantly damped due to momentum scattering.
Its influence is observable in the strong temperature dependence of the decay constant, which is reduced from 35~ps to 18~ps as the temperature is increased from 1.3~K to 4.5~K due to the reduced electron mobility, and the increasing importance of electron-electron scattering (Fig. \ref{Zerofield}(b)). In this sample, the ensemble momentum scattering time $\tau_p$, which may be extracted from mobility measurements, and the electron-electron scattering time (see Table~\ref{Data}), are comparable at 4~K.

The situation is drastically different in a symmetric $(110)$-grown QW, sample C (Fig. \ref{Zerofield}(c)). Here, the Dresselhaus field points along the growth direction and is therefore parallel to optically oriented electron spins. Due to the symmetric doping, there is no pronounced Rashba field present in this sample, therefore, the
injected electron spins do not precess coherently. The partial decay of the Faraday signal in this sample is due to photocarrier recombination and the BAP mechanism.~\cite{bap} In symmetrically-grown $(110)$ samples, very long spin dephasing times have been
observed, and other dephasing mechanisms may be important.~\cite{Dohrmann,Muller09} The limit for spin dephasing in these systems is due to random Rashba fields caused by fluctuations in the remote doping density.~\cite{Sherman_Wu}

We now turn to measurements in weak perpendicular magnetic fields and focus on sample A.
\begin{figure}
  \includegraphics[width= 0.3 \textwidth]{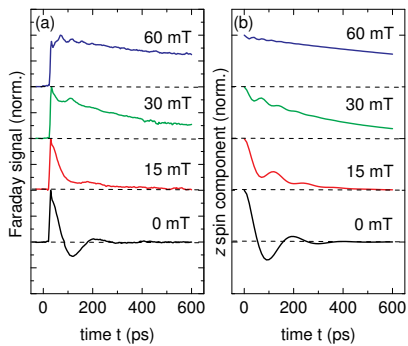}
   \caption{(a) TRFR measurements in various small perpendicular  magnetic fields measured on sample A at 400~mK.
   (b) Calculated time evolution of the $z$-component of the spin polarization for magnetic fields corresponding to measurements in (a).}
   \label{Weakfield}
\end{figure}
Figure \ref{Weakfield} (a) shows a series of TRFR traces measured on sample A for various perpendicular magnetic fields at 400~mK. Two features are clearly visible: at $B>0$,  fast, damped spin beats  become apparent in the Faraday signal, their frequency increases with the perpendicular magnetic field. The amplitude of these beats is small, and for $B>10$~mT, the Faraday signal does not cross the zero line. Additionally, the decay of the Faraday signal becomes slower with increasing magnetic field, and after the spin beats are damped out, a longer-lived tail of the Faraday signal is observed.

\begin{figure}
  \includegraphics[width= 0.4 \textwidth]{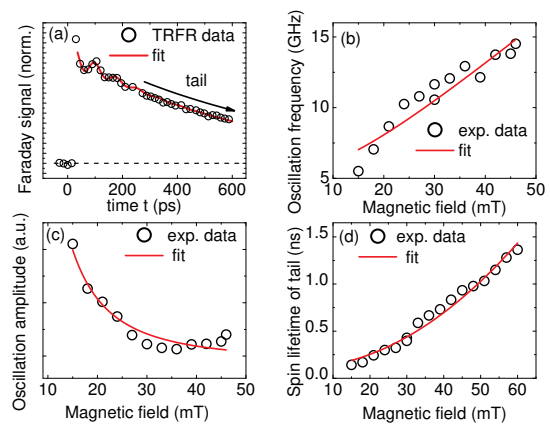}
   \caption{(a) TRFR  measured on Sample A at 400~mK in a 36~mT perpendicular magnetic field (open circles). The fit function is shown as a red solid line. The arrow in the figure traces the `tail' of the spin polarization decay.
   (b) Oscillation frequency of the  spin beats observed in perpendicular magnetic fields. The red solid line is a fit to the experimental data (open circles). (c) Amplitude of the spin beats as a function of magnetic field. The red solid line is a fit to the experimental data. (d) Spin lifetime of the `tail' of the spin polarization decay as a function of perpendicular magnetic field. The red solid line is a fit to the experimental data (open circles) using the approximate Eq.~\eqref{szt}.}
   \label{Weakfield_Fit}
\end{figure}

The quantitative description of the electron spin dynamics in a magnetic field is carried out within the standard kinetic approach.~\cite{Glazov_SSC,gi} The kinetic equation for the spin distribution function $\bm s_{\bm k}$ in a
magnetic field $\bm B$ applied along the growth axis can be written as $%
\partial \bm s_{\bm k}/\partial t+\bm s_{\bm k}\times \bm \Omega _{\bm %
k}+\omega _{c}\partial \bm s_{\bm k}/\partial \varphi _{\bm k}+Q\{\bm s_{\bm %
k}\}=0$. Here  $\omega _{c}=\left| e\right| B/mc$ is the cyclotron
frequency, where $e$ and $m$ are the electron charge and effective mass,
respectively, $ \varphi _{\bm k}$ is the angle between ${\bm k}$
and the $x-$axis, and $\bm Q\{\bm s_{\bm k}\}$ is the collision integral. The main features of the spin dynamics can be understood from the solution of the kinetic equation in the case of isotropic spin splitting $|\bm \Omega_{\bm k}| = \Omega_{k}$ under the assumption $\Omega_{k_{\rm F}}\tau^*,\omega_{c}\tau^*\gg 1$:~\cite{assump}
\begin{equation}
\label{szt}
\frac{s_{z,k_{\mathrm F}}(t)}{s_{z,k_{\mathrm F}}(0)} = \mathcal A \mathrm e^{-t/\tau_s} + \mathcal B \mathrm e^{-t/\tau_b}\cos{(\Omega_{\rm eff} t)},
\end{equation}
where $\Omega_{\rm eff} = \sqrt{\omega_{c}^2 + \Omega_{k_{\rm F}}^2}$, $\mathcal A = \omega_{c}^2/\Omega_{\rm eff}^2$, $\mathcal B = \Omega_{k_{\rm F}}^2/\Omega_{\rm eff}^2$, $\tau_s^{-1} = \mathcal B/\tau^*$, ${\tau_b}^{-1}=(1+\mathcal A)/(2\tau^*)$. In the  general case of anisotropic spin splitting and non-zero temperature Eq.~\eqref{szt} approximately holds for the averaged spin if $\Omega_{k_{\rm F}}$ and $1/\tau^*$ are replaced by the effective values containing an additional contribution due to the spread of spin precession frequencies.~\cite{Glazov_SSC} According to Eq.~\eqref{szt}, the $z$-component of the spin, $s_{z,k_{\mathrm F}}(t)$, demonstrates damped oscillations with the combined frequency $\Omega_{\rm eff}$, superimposed on the exponential decay. The enhancement of the spin precession frequency is a result of the electron cyclotron motion. The cyclotron rotation of $\bm k$ results in the modulation of the SO field and the electron spin experiences a torque being a geometrical sum of $\bm \Omega_{k_{\rm F}}$ and $\bm \omega_c$.~\cite{Glazov_SSC}

The damping of the spin beats occurs at time scale $\tau_b\sim \tau^*$, while the decay of the average spin value takes a much longer time $\tau_s\gg \tau^*$. At $\Omega_{k_{\rm F}} \ll \omega_{c}$, the spin relaxation time $\tau_s$ passes to the `motional narrowing' expressions~\cite{Ivchenko73,glazov04} because the variation of the SO field caused by the cyclotron rotation is much faster than the spin precession in the field $\bm \Omega_{k_{\rm F}}$.  The results of a numerical solution of the kinetic equation, with allowance for anisotropic spin splitting,~\cite{model_param} are shown in Fig.~\ref{Weakfield}(b). The experimental data are well reproduced by such a modeling.

For further comparison between experiment and theory we plot the amplitude and frequency of the fast spin beats, as well as the decay constant of the tail, by fitting the experimental data by Eq.~\eqref{szt}, as it is shown in Fig.~\ref{Weakfield_Fit} (a). The summary of the fit parameters is given in panels (b)-(d) of Fig.~\ref{Weakfield_Fit}. In Fig.~\ref{Weakfield_Fit} (b), we plot the observed frequency as a function of the applied magnetic field. The best fit using the geometric sum formula is shown as a solid red line. It is in good agreement with the experimental data, however,  the effective cyclotron frequency extracted from the fit is about $20\%$ larger as compared to the calculated one. Such a difference can be caused by the anisotropy of the spin splitting. The beats at low magnetic field are damped on the same timescale as the coherent zero-field oscillation in agreement with Eq.~\eqref{szt} due to the anisotropy of the SO field and the momentum randomization due to disorder and electron-electron scattering.~\cite{Leyland_ee,gi}

The spin beats amplitude systematically decreases with the applied magnetic field, as shown in  Fig.~\ref{Weakfield_Fit}(c), because fast cyclotron rotation of $\bm \Omega_{\bm k}$ suppresses the spin precession around $\bm \Omega_{\bm k}$. The red solid line in Fig. \ref{Weakfield_Fit} (c), which represents a fit to the data using Eq.~\eqref{szt}, is in good agreement with the experimental data.

Finally, we focus on the magnetic field dependence of the long-lived `tail' of the Faraday signal. The experimental data in Fig. \ref{Weakfield_Fit}(d) demonstrate that the decay time of this tail increases quadratically as a function of the perpendicular field. A fit according to Eq.~\eqref{szt} (solid red line in Fig. \ref{Weakfield_Fit})  leads to a rather small scattering time $\tau^*\approx 20$~ps, which is an evidence of the strong spin splitting anisotropy in the sample. The kinetic model with allowance for the spin splitting anisotropy~\cite{model_param} (not shown in the Figure) is in  excellent agreement with the experimental findings.

In the symmetric [110]-grown QW, where the SO field points out of the QW plane,  this precession is absent, as the optically oriented electron spins are parallel to the externally applied magnetic field and the SO field. TRFR measurements on sample C at low temperatures in weak perpendicular fields (not shown)  show no  spin beats, confirming that  there is no significant Rashba field in this sample.

In conclusion, we have investigated the electron spin dynamics in high-mobility two-dimensional electron systems grown in different crystallographic directions at low temperatures. In measurements without applied magnetic field,  pronounced coherent oscillations of the electron spin polarization about the SO field are observed in the $(001)$-grown samples. In the same samples
non-quantizing perpendicular magnetic fields make the spin beats faster and strongly suppress the decay of the net spin polarization due to the cyclotron rotation of the effective SO field. In symmetric $(110)$-grown samples the SO field does not result in a $z$ spin component rotation and, consequently, no cyclotron effect is observed. The experimental observations are in good agreement with calculations based on a kinetic equation approach.

The authors would like to thank E.L. Ivchenko for very fruitful discussions. Financial support by the DFG via SPP 1285 and SFB 689,
University of Basque Country Grant GIU07/40, RFBR, Dynasty Foundation and President grant for young scientists is gratefully acknowledged.


\begin{thebibliography}{99}
\bibitem{assump}
For the parameters used in calculations~\cite{model_param} these quantities are: $\Omega_{k_{\rm F}} \tau^* \approx 2.6$, $\omega_c(10\mbox{ mT})\tau^* \approx 2.5$.
\bibitem{model_param}
The scattering time $\tau^*=88$~ps, the Dresselhaus SO field $\bm \Omega_{\bm k} =[-\Omega_1\cos{\varphi}-\Omega_3\cos{3\varphi},\Omega_1\sin{\varphi}-\Omega_3\sin{3\varphi}]$, $\varphi$ is the angle between $\bm k$ and $x\parallel[100]$, $\Omega_1=\Omega_3 =0.03$~ps$^{-1}$, the scattering potential is taken to be smooth.
\end{thebibliography}
\end{document}